\documentclass{emulateapj}
\usepackage{amsmath}
\usepackage{ulem}
\shorttitle{FRB/GRB cosmology} \shortauthors{Deng \& Zhang} \slugcomment{}
\begin{document}
\title{Cosmological implications of Fast Radio Burst / Gamma-Ray Burst Associations}
\author{Wei Deng, Bing Zhang}
\affil{Department of Physics and Astronomy, University of Nevada Las Vegas, Las Vegas, NV 89154, USA\\deng@physics.unlv.edu, zhang@physics.unlv.edu}

\begin{abstract}
If a small fraction of Fast Radio Bursts (FRBs) are associated with Gamma-Ray Bursts (GRBs), as recently suggested by Zhang, the combination of redshift measurements of GRBs and dispersion measure (DM) measurements of FRBs opens a new window to study cosmology. At $z<2$ where the universe is essentially fully ionized, detections of FRB/GRB pairs can give an independent measurement of the intergalactic medium portion of the baryon mass fraction, $\Omega_b f_{\rm IGM}$, of the universe. If a good sample of FRB/GRB associations are discovered at higher redshifts, the free electron column density history can be mapped, which can be used to probe the reionization history of both hydrogen and helium in the universe. We apply our formulation to GRBs 101011A and 100704A that each might have an associated FRB, and constrained $\Omega_b f_{\rm IGM}$ to be consistent with the value derived from other methods. The methodology developed here is also applicable, if the redshifts of FRBs not associated with GRBs can be measured by other means.
\end{abstract}

\keywords{gamma-rays: bursts - cosmology: cosmological parameters, reionization - radio: bursts }

\section{Introduction\label{sec:intro}}

The physical origin of newly discovered fast radio bursts \citep[FRBs,][]{lorimer07,thornton13} is debated \citep{thornton13,falcke13,totani13,popov07,popov13,kashiyama13,loeb13,zhang14,kulkarni14}. One attractive proposal is delayed collapses of supra-massive neutron stars after loosing centrifugal support due to spin down \citep{falcke13}. \cite{zhang14} recently suggested that within such a scenario, a small fraction of FRBs can be physically associated with some gamma-ray bursts (GRBs), whose central engine is a supra-massive millisecond magnetar, which collapses into a black hole after the GRB prompt emission is over ($10^2-10^4$ s). Such a FRB/GRB association might have been detected in GRB 101011A and GRB 100704A by \cite{bannister12}. If such FRB/GRB associations are confirmed to be common, it opens a new window to study cosmology\footnote{\cite{ioka03} has discussed measuring DM of a GRB using radio afterglows. However, lacking a clear impulsive radio emission signal, such a measurement is difficult to realize. The FRB/GRB associations are ideal systems to achieve such a goal.}. This Letter discusses the cosmological implications of such associations.

\section{Dispersion measure of FRB/GRB systems}\label{sec:overview}

For an FRB/GRB association system, one can in principle get two precise measurements. One is the redshift of the system, which can be measured from the emission lines of the GRB host galaxies or absorption lines of the GRB afterglows. The second is the dispersion measure (DM) of the system measured from the FRB. In general, the DM is defined as the delayed arrival time of a radio wave with respect to the arrival time in vacuum, i.e. \citep{rybicki79}
\begin{equation}
\Delta t \simeq \int \frac{dl}{c} \frac{\nu_p^2}{2\nu^2} \simeq 4.2 ~{\rm s}~ \left(\frac{\nu}{1~{\rm GHz}}\right)^{-2} \frac{\rm DM}{10^3~{\rm pc~cm^{-3}}},
\end{equation}
where $\nu_p = (ne^2/\pi m_e)^{1/2} = 8.98 \times 10^3 n_e^{1/2}$ Hz is the plasma frequency, and DM is normalized to a typical value $10^3~{\rm pc~cm^{-3}}$ for the intergalactic medium (IGM) to a source at a cosmological distance. Practically it is measured from the time delay between two frequencies. For a plasma at redshift $z$, the rest-frame delay time ($\Delta t_{\rm z}$) between two rest-frame frequencies ($\nu_{1,\rm z}<\nu_{2,\rm z}$) is
\begin{eqnarray}
\Delta t_{\rm z} & = & \int \frac{dl}{c} \frac{\nu_{\rm p}^2}{2} \left(\frac{1}{\nu_{1,\rm z}^2}-\frac{1}{\nu_{2,\rm z}^2}\right) \nonumber \\
& = & \frac{e^2}{2\pi m_{\rm e} c} \left(\frac{1}{\nu_{1,\rm z}^2}-\frac{1}{\nu_{2,\rm z}^2}\right)\int n_{\rm e,z}dl,
\label{eq:Delta_t_z}
\end{eqnarray}
where $\int n_{\rm e,z}dl = {\rm DM}_z$ is the rest-frame dispersion measure, which is just the column density of free electrons at the source. In the observer frame, the observed delay time is $\Delta t=\Delta t_{\rm z}\times (1+z)$ and the observed frequency is $\nu=\nu_{\rm z}/(1+z)$. So Eq.(\ref{eq:Delta_t_z}) can be modified as
\begin{equation}
\Delta t = \frac{e^2}{2\pi m_{\rm e} c} \left(\frac{1}{\nu_{1}^2}-\frac{1}{\nu_{2}^2}\right)\int \frac {n_{\rm e,z}}{1+z}dl,
\label{eq:Delta_t}
\end{equation}
where the measured DM by an earth observer is
\begin{equation}
{\rm DM}=\int \frac {n_{\rm e,z}}{1+z}dl.
\label{eq:DMz}
\end{equation}

For an FRB/GRB system, the measured DM should include four terms:
\begin{equation}
{\rm DM}_{\rm tot}={\rm DM}_{\rm MW}+{\rm DM}_{\rm IGM}+{\rm DM}_{\rm host}+{\rm DM}_{\rm GRB}.
\label{eq:DM}
\end{equation}
They denote dispersion measure contributions from the Milky Way, intergalactic medium, GRB host galaxy, and the GRB blastwave itself, respectively. The observed ${\rm DM}_{\rm tot}$ of FRBs are around several hundreds ${\rm pc\,cm^{-3}}$ \citep{lorimer07,thornton13}, and the two putative FRBs associated with two GRBs also have similar values of DMs \citep{bannister12}. In the following we discuss the relative importance of the four terms in turn.

\subsection{${\rm DM}_{\rm MW}$ and ${\rm DM}_{\rm host}$}

${\rm DM}_{\rm MW}$ is well constrained with the pulsar data \citep{taylorcordes93}, and is a strong decreasing function of Galactic latitude $|b|$, from ${\rm DM}^{\rm max}_{\rm MW}\sim 10^3~{\rm pc\,cm^{-3}}$ when $|b|\sim0^{\rm o}$ to $<100~{\rm pc\,cm^{-3}}$ at $|b| > 10^{\rm o}$ \citep{thornton13}. The observed FRBs all have relatively large $|b|$, so ${\rm DM}_{\rm MW}$ is a relatively small term.

${\rm DM}_{\rm host}$ is poorly known. If GRBs are born in giant molecular clouds, ${\rm DM}_{\rm host}$ may be very large \citep{ioka03}. Afterglow studies of GRBs seem to suggest that the GRB circumburst density is relatively low, with a typical value of $n_{\rm ISM} \sim 1~{\rm cm^{-3}}$ \citep[e.g.][]{panaitescu02,yost03}. Considering that the GRB host galaxies are typically smaller than Milky Way \citep{fruchter06} and the additional $(1+z)$ deduction factor (Eq.(\ref{eq:DMz})), it would be reasonable to assume that on average ${\rm DM}_{\rm host} \leq {\rm DM}_{\rm MW}$\footnote{It is possible that in a small fraction of FRB-GRB association systems, $DM_{\rm host}$ may be anomalously high, probably due to the existence of a dense thick electronic disk viewed near the edge-on direction. Such outliers can be easily recognized, and can be excluded for cosmological studies discussed in this paper.}.

\subsection{${\rm DM}_{\rm GRB}$}\label{sec:DM_GRB}

A GRB-associated FRB would happen at the end of the X-ray plateau phase (or somewhat later), which could be the time when the supramassive neutron star collapses into a black hole \citep{zhang14}\footnote{For an internal plateau, the FRB time is supposed to be the beginning of the steep decay phase. For a normal plateau, the supra-massive neutron star can collapse at the end of plateau, or any other time after the plateau, depending on the mass of the neutron star and equation of state of the nuclear matter.}. As the FRB is ejected at an inner radius $r_1 \sim 10^7$ cm, the blastwave is already at a large radius $r_2 \sim c \delta t = 3\times 10^{13}~{\rm cm}{\left(\frac{\delta t}{1000~{\rm s}}\right)}$, where $\delta t$ is the delay time between the FRB and the GRB. The FRB, traveling essentially with speed of light, would catch up with the blastwave at a radius $r_3 \sim 2[\Gamma(r_3)]^2 r_2 \gg r_2$. At this radius, the plasma frequency is much lower than the FRB frequency, so that the FRB can go through the blastwave \citep{zhang14}. In any case, the blastwave would contribute to the frequency dispersion, which we calculate below.

One important parameter is the baryon loading parameter of the GRB, which may be characterized as $\Gamma_0 = E_{\rm iso} / M_0 c^2$, where $M_0$ is the initial mass loading in the GRB outflow, and $E_{\rm iso} = E_{\rm \gamma,iso} + E_{\rm X,iso} + E_{\rm K,iso}$ is the isotropic energy of the GRB, which is the sum of the isotropic energy released in $\gamma$-rays (prompt phase), in X-rays as internal emission (during the internal plateau), and the isotropic kinetic energy that powers the afterglow emission (the normal plateau) \citep{lv14}. For FRB-associated GRBs, there should be energy injection in the early afterglow phase \citep{zhang14}, so the kinetic energy $E_{\rm K,iso}$ should be calculated after energy injection is over. The parameter $\Gamma_0$ is therefore the final ``effective'' initial Lorentz factor of the outflow. It reflects the average baryon-loading parameter $\eta (1+\sigma_0)$ ($\eta$ is the dimensionless entropy, and $\sigma_0$ is the initial magnetization parameter at the central engine) \citep{lei13}. The radius at which the FRB catches up the blastwave also depends on the density profile of the circumburst medium, which could be either a constant density medium with $\rho = n m_{\rm p}$ ($n$ is the number density of protons/electrons) or a stellar wind with $\rho =Ar^{-2}$ ($A=5\times 10^{11} A_\star ~{\rm g~cm}^{-1}$ is the wind parameter).

At late phase of blastwave propagation (much later than the energy injection phase), which is relevant for FRB catching up with the blastwave, the energy conservation equation can be written as
\begin{equation}
\Gamma_0 M_0 + m(r)=\Gamma(r)[M_0 + \Gamma(r) m(r)],
\label{eq:Gamma}
\end{equation}
where $m(r)$ is the mass accumulated from the circumburst medium, which is $m(r)=(4/3)\pi(r^3-r_1^3)n m_{\rm p}$ for ISM, and $m(r)=\int^r_{r_1} \rho \pi r^2 dr=\pi A (r-r_1)$ for wind.

The catching up condition can be more rigorously written as
\begin{equation}
\frac{r_3-r_1}{c}=\int_{r_2}^{r_3}\frac{dr}{\beta(r)c}.
\label{eq:r3}
\end{equation}
Using Eqs. (\ref{eq:Gamma}) and (\ref{eq:r3}), one can solve for $r_3$ for different initial parameters. Since the thickness $\Delta$ of the blastwave is $\ll r_3$, one can calculate ${\rm DM}_{\rm GRB}$ in the observer frame as
\begin{eqnarray}
{\rm DM}_{\rm GRB}&=&\frac{{\rm DM}_{\rm GRB,z}}{1+z}= \frac{\int n_{\rm e} dl}{1+z}\nonumber\\
&\simeq&  \frac{[M_0+m(r_3)]/m_{\rm p}}{(1+z) \pi r_3^2 \Delta} \Delta =  \frac{M_0+m(r_3)}{(1+z) m_p \pi r_3^2}.
\label{eq:DM_GRB}
\end{eqnarray}

Based on the above equations, we calculate the rest frame GRB DM value, ${\rm DM}_{\rm GRB,z}$ for a set of typical values of GRB parameters \citep{panaitescu02,yost03,zhang04}: $r_1=10^7~{\rm cm}$, $E_{\rm iso}=10^{53}~{\rm erg}$, $\Gamma_0=300$ (so that $M_0=3.7\times10^{29}~{\rm g}$), $\delta t=500~{\rm s}$, $n=1~{\rm cm}^{-3}$ (ISM) and $A_\star=1$ (wind). For the ISM case, we get $r_3=2.0\times10^{17}~{\rm cm}$, $\Gamma(r_3)=42.7$, $m(r_3)=5.2\times10^{28}~{\rm g}$, and ${\rm DM}_{\rm GRB,z}=0.68~{\rm pc\,cm^{-3}}$; For the wind case, we get $r_3=6.0\times10^{16}~{\rm cm}$, $\Gamma(r_3)=32.4$, $m(r_3)=9.5\times10^{28}~{\rm g}$, and ${\rm DM}_{\rm GRB,z}=7.9~{\rm pc\,cm^{-3}}$. These are summarized in the first row of Table \ref{table:DM_GRB}. In the following rows in Table \ref{table:DM_GRB}, we vary each input parameter ($E_{\rm iso}$, $\Gamma_0$, $\delta t$, and $n/A_*$) to a wider range and recalculate the ${\rm DM}_{\rm GRB,z}$ values. In particular, we incorporate more extreme parameters \citep{zhang04} in favor of large ${\rm DM}_{\rm GRB}$ values. The general result is that ${\rm DM}_{\rm GRB,z}$ is much less than ${\rm DM}_{\rm tot}$ detected from FRBs, and in most cases even smaller than ${\rm DM}_{\rm MW}$ and ${\rm DM}_{\rm host}$. Correcting for the $(1+z)$ factor (Eq.(\ref{eq:DMz})), the value is even smaller. For parameter dependences, ${\rm DM}_{\rm GRB,z}$ is larger for a larger mass loading (smaller $\Gamma_0$), shorter time delay $\delta t$ (so that the FRB catches up with the blastwave at a smaller radius), or a higher ambient density $n$ or $A_*$ (again a smaller catch-up radius). The effect of isotropic energy $E_{\rm iso}$ is mixed: it increases baryon loading (given a same $\Gamma_0$) but also increases catch-up radius by increasing deceleration radius. So ${\rm DM}_{\rm GRB}$ tends to increase for ISM but decrease for wind when $E_{\rm iso}$ increases. Unless extreme parameters in a wind medium is invoked, ${\rm DM}_{\rm GRB}$ is negligible in Eq.(\ref{eq:DM}).

\begin{table}
\caption{The calculated ${\rm DM}_{\rm GRB,z}$ with different parameters.}
\begin{tabular}{|l|l|l|l|l|}
\hline & \multicolumn{2}{|c|}{ISM} & \multicolumn{2}{|c|}{wind}\\\hline
${\rm DM}_{\rm GRB,z}$(typical) & \multicolumn{2}{|c|}{0.68} & \multicolumn{2}{|c|}{7.9}\\\hline
${\rm DM}_{\rm GRB,z}$($E_{\rm iso}$)& 0.23($10^{52}$)& 2.1($10^{54}$) &10.7($10^{52}$)& 7.7($10^{54}$)\\\hline
${\rm DM}_{\rm GRB,z}$($\Gamma_0$)& 2.9(100) & 0.36(600) & 28.7(100) & 4.4(600)\\\hline
${\rm DM}_{\rm GRB,z}$($\delta t$)& 1.6(100) & 0.5(1000) & 37.7(100) & 4.2(1000)\\\hline
${\rm DM}_{\rm GRB,z}$($n/A_\star$)& 0.21(0.1) & 2.3(10) & 0.77(0.1) & 107(10)\\\hline
\end{tabular}
\label{table:DM_GRB}\\ \\
\renewcommand{\thefootnote}
\thefootnote{The units of the parameters: ${\rm DM}_{\rm GRB,z}$ in ${\rm pc~cm^{-3}}$; $E_{\rm iso}$ in ${\rm erg}$; $\delta t$ in ${\rm s}$; $n$ in ${\rm cm^{-3}}$. ${\rm DM}_{\rm GRB,z}$ (typical) is the value of ${\rm DM}_{\rm GRB,z}$ with typical parameters introduced in section \S\ref{sec:DM_GRB}. The following four rows present the calculated values of ${\rm DM}_{\rm GRB,z}$ by changing one parameter (in parenthesis) with other parameters kept as the typical values.}
\end{table}

For a quick estimate, one can also use an approximated treatment to derive $r_3$ and ${\rm DM}_{\rm GRB,z}$. Since the catch-up radius is still in the relativistic phase of the blastwave, one can simplify Eq.(\ref{eq:Gamma}) to $\Gamma_0 M_0=\Gamma(r)^2 m(r)$, or $E_{\rm iso} = \Gamma(r_3)^2 m(r_3) c^2$. Noticing the catch up condition $r_3 \simeq 2 \Gamma(r_3)^2 r_2 = 2 \Gamma(r_3)^2 c \delta t$, one can derive
\begin{equation}
r_3 = \left(\frac{3E\delta t}{2 \pi n m_p c}\right)^{1/4} \simeq 1.5\times 10^{17}~{\rm cm}E_{\rm iso,53}^{1/4}
\left(\frac{\delta t}{500~{\rm s}}\right)^{1/4} n^{-1/4}
\end{equation}
for ISM, and
\begin{equation}
r_3 = \left(\frac{2E\delta t}{\pi A c}\right)^{1/2} \simeq 4.6\times 10^{16}~{\rm cm}E_{\rm iso,53}^{1/2}
\left(\frac{\delta t}{500~{\rm s}}\right)^{1/2} A_\star^{-1/2}
\end{equation}
for wind. These are about 25\% smaller than the numerical values ($2.0\times 10^{17}$ cm, and $6.0 \times 10^{16}$ cm, respectively), which leads to over-estimate of ${\rm DM}_{\rm GRB,z}$ by about 60\% (using Eq.(\ref{eq:DM_GRB})).

\subsection{${\rm DM}_{\rm IGM}$}

The largest contribution to ${\rm DM}_{\rm tot}$ (Eq.(\ref{eq:DM})) is from the ionized IGM, i.e. ${\rm DM}_{\rm IGM}=\int \frac {n_{\rm e}}{1+z}dl$. \cite{ioka03} and \cite{inoue04} have derived some equations of ${\rm DM}_{\rm IGM}$ for a fully ionized, pure hydrogen plasma. Here we derive a more general expression. We consider an IGM with a hydrogen (H) mass fraction $Y_{\rm H} = (3/4) y_1$ and helium (He) mass fraction $Y_{\rm He} = (1/4) y_2$, where $y_1 \sim 1$ and $y_2 \simeq 4 - 3 y_1 \sim 1$ are the hydrogen and helium mass fractions normalized to the typical values 3/4 and 1/4, respectively. We also introduce the ionization fractions for each species as a function of redshift\footnote{The two electrons of He have different ionization energies. The parameter $\chi_{\rm e,He}$ is a weighted ionization fraction of the two electrons.}, i.e. $\chi_{\rm e,H}(z)$ and $\chi_{\rm e,He}(z)$. The number density of free electrons at redshift $z$ can be expressed as
\begin{eqnarray}
n_{\rm e}&=&n_{\rm H,0}(1+z)^3\,\chi_{\rm e,H}(z)+ 2\, n_{\rm He,0}(1+z)^3\, \chi_{\rm e,He}(z) \nonumber \\
&=&\left[\frac{Y_{\rm H}~ \rho_{\rm c, 0}\Omega_{\rm b}f_{\rm IGM}}{m_{p}}\,\chi_{\rm e,H}(z) + 2\frac{Y_{\rm He}~\rho_{\rm c, 0}\Omega_{\rm b}f_{\rm IGM}}{4 m_{p}}\, \chi_{\rm e,He}(z)\right] \nonumber \\
& & \times  (1+z)^3 \nonumber \\
&=&\frac{\rho_{\rm c, 0}\Omega_{\rm b}f_{\rm IGM}}{m_{p}}\,\left[\frac{3}{4} y_1 \chi_{\rm e,H}(z)+\frac{1}{8} y_2 \chi_{\rm e,He}(z)\right](1+z)^3.
\label{eq:ne}
\end{eqnarray}
Here $n_{\rm H,0}$ and $n_{\rm He,0}$ are the number density of H and He at $z=0$, $\rho_{c,0}$ is the critical mass density at $z=0$, $\Omega_b$ is the current baryon mass fraction of the universe, and $f_{\rm IGM}$ is the fraction of baryon mass in the intergalactic medium. Noticing
\begin{equation}
dl=\frac{1}{1+z}\frac{c}{H_0}\frac{dz}{\sqrt{\Omega_m(1+z)^3+\Omega_\Lambda}}
\end{equation}
for a flat ($k=0$) universe, one gets
\begin{eqnarray}
{\rm DM}_{\rm IGM} & = & \frac{3cH_0\Omega_bf_{\rm IGM}}{8\pi G m_p} \nonumber \\
& \times & \int_{0}^{\rm z} \frac{[\frac{3}{4} y_1 \chi_{\rm e,H}(z)+\frac{1}{8} y_2 \chi_{\rm e,He}(z)](1+z)dz}{[\Omega_m(1+z)^3+\Omega_\Lambda]^{1/2}}.
\label{eq:DM_IGM}
\end{eqnarray}

\section{Measure $\Omega_b f_{\rm IGM}$ with FRB/GRB systems}\label{sec:Omega_b}

The baryon mass fraction $\Omega_b$ is an important parameter in cosmology. Currently it is measured through Big Bang nucleosynthesis \citep{walker91} or anisotropy data of cosmic microwave background \citep{WMAP9yrs,planckcollaboration13}.
The derived results vary from 0.02 to 0.05. The latest Planck + WMAP results \citep{planckcollaboration13} give $\Omega_b = (0.0458,0.0517)$ within $2\sigma$.

The FRB/GRB systems provide an independent method to directly measure the IGM portion of baryon mass fraction, $\Omega_b f_{\rm IGM}$. Re-writing Eq.(\ref{eq:DM_IGM}), one gets
\begin{eqnarray}
\Omega_b f_{\rm IGM} & = & \frac{8\pi G m_p {\rm DM}_{\rm IGM}}{3cH_0} \nonumber \\
& / &  \int_{0}^{\rm z} \frac{[\frac{3}{4} y_1 \chi_{\rm e,H}(z)+\frac{1}{8} y_2 \chi_{\rm e,He}(z)](1+z)dz}{[\Omega_m(1+z)^3+\Omega_\Lambda]^{1/2}}
\label{eq:Omega_b_full}
\end{eqnarray}
Studies suggest that H is essentially fully ionized at $z<6$ \citep{fanx06}, and He is essentially fully ionized at $z<2$ \citep{mcquinn09}. So for nearby GRBs ($z<2$), one can take $\chi_{\rm e,H} = \chi_{\rm e,He} = 1$. When taking $y_1 \sim y_2 \sim 1$, one has
\begin{eqnarray}
\Omega_b f_{\rm IGM} \simeq  \frac{64\pi G m_p {\rm DM}_{\rm IGM}}{21 cH_0} / \int_{0}^{\rm z} \frac{(1+z)dz}{[\Omega_m(1+z)^3+\Omega_\Lambda]^{1/2}} \nonumber \\
\label{eq:Omega_b}
\end{eqnarray}
Since $H_0$, $\Omega_m$ and $\Omega_\Lambda$ can be well measured by other methods, by measuring $z$ and ${\rm DM}_{\rm IGM}$ of FRB/GRB systems at $z<2$, one can directly measure $\Omega_b f_{\rm IGM}$.

Two GRBs, 101011A and 100704A, might each have an associated FRB \citep{bannister12} with properties similar to other FRBs \citep{lorimer07,thornton13}\footnote{ The significance of the signals was low, and \cite{bannister12} were not certain about whether the associations are real. On the other hand, the epochs of FRBs are consistent with the theoretically motivated epochs as discussed in \cite{zhang14}, which suggests that these two FRBs may be real.}. Unfortunately, neither GRB had a measured redshift, so that our method cannot be applied directly. Nonetheless, we can apply some empirical relations to estimate the redshift range of the two GRBs, and hence, pose a constraint on $\Omega_b f_{\rm IGM}$.

\begin{figure}
\plotone{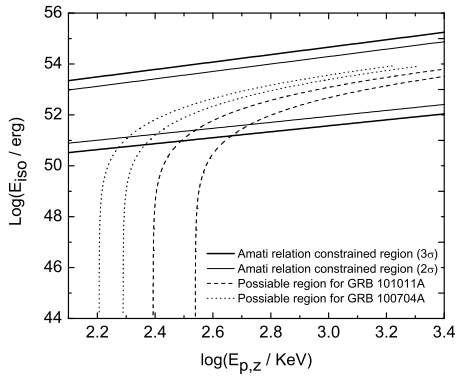}
\caption{Apply Amati relation to constrain the possible redshift range of GRB 101011A and GRB 100704A. The thick (thin) solid lines enclose the $3\sigma$ ($2\sigma$) regions of the correlation. The upper $3\sigma$ line is defined by adopting $\gamma=1.46+0.29$, ${\rm A}=49.17+3\sigma_{\rm ext}$, while the lower $3\sigma$ line is defined by adopting $\gamma=1.46-0.29$, ${\rm A}=49.17-3\sigma_{\rm ext}$. The $2\sigma$ region is defined similarly. The region between two dashed lines and two dotted lines are the possible positions for GRB 101011A and GRB 100704A on this plot with different redshifts. The requirement that the GRBs enter the 3$\sigma$ region gives $z \geq 0.246$ for GRB 101011A and $z \geq 0.166$ for GRB 100704A.}
\label{fig:redshift}
\end{figure}

We apply the so-called Amati relation \citep{amati02,amati08,capozziello10}
\begin{equation}
\log\frac{E_{\rm \gamma,iso}}{\rm erg}=A+\gamma \log\frac{E_{\rm p,z}}{\rm keV}
\label{eq:Amati}
\end{equation}
to estimate the redshifts of the two GRBs. Here $E_{\rm \gamma,iso}$ is the normalized isotropic $\gamma$-ray energy of the GRB, $E_{\rm p,z}$ is the intrinsic peak energy with redshift correction, and the fitting parameters are $A=49.17\pm0.40, \gamma=1.46\pm0.29$, with a standard deviation $\sigma_{\rm ext}=0.37$ \citep{capozziello10}. In Fig.\ref{fig:redshift} we draw $3\sigma$ and $2\sigma$ zones of the correlation, and use the observed fluence and $E_{\rm p}$ of the two bursts (GRB 101011A \citep{burgess10}: 8-1000 keV fluence $(5.24\pm 0.39) \times 10^{-6}~{\rm erg~cm^{-2}}$ and $E_p = 296.6 \pm 49.4$ keV; GRB 100704A \citep{mcbreen10}: 10-1000 keV fluence $(5.8\pm 0.2) \times 10^{-6} ~{\rm erg~cm^{-2}}$ and $E_p = 178.30^{+16.30}_{-17.50}$ keV) to calculate the intrinsic $E_{\rm iso}$ and $E_{\rm p,z}$ for different redshifts. For each burst, we draw two curves to reflect the errors of the observables. By requiring that the bursts enter the $3\sigma$ region of the correlation, we derive $z \geq 0.246$ for GRB 101011A and $z \geq 0.166$ for GRB 100704A.

With the constrained redshift range, we can then constrain the range of $\Omega_b f_{\rm IGM}$ using Eq.(\ref{eq:Omega_b}). The standard cosmological parameters derived by the latest Planck team \citep{planckcollaboration13} are adopted $(\Omega_{\rm m}, \Omega_{\rm \Lambda}, h)=(0.315, 0.685, 0.673)$: The measured ${\rm DM}_{\rm tot}$ values are $569.98~{\rm pc~cm^{-3}}$ for GRB 101011A, and $194.57~{\rm pc~cm^{-3}}$ for GRB 100704A \citep{bannister12}. These are the upper limits of ${\rm DM}_{\rm IGM}$. According to \cite{taylorcordes93} and \cite{thornton13}, ${\rm DM}_{\rm MW}$ of the two GRBs would be about $30~{\rm pc\,cm^{-3}}$ for GRB 101011A ($|b|=45.4^{\rm o}$) and $40~{\rm pc\,cm^{-3}}$ for GRB 100704A ($|b|=13.2^{\rm o}$). For simplicity we assume ${\rm DM}_{\rm host}={\rm DM}_{\rm MW}$, and neglect ${\rm DM}_{\rm GRB}$. We then get possible values of ${\rm DM}_{\rm IGM}$: about $510~{\rm pc\,cm^{-3}}$ for GRB 101011A and $115~{\rm pc\,cm^{-3}}$ for GRB 100704A. %

\begin{figure}[ht]
\plotone{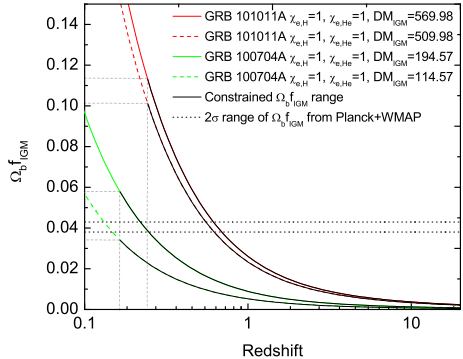}
\caption{Constraints on $\Omega_{b} f_{\rm IGM}$ using the two FRB/GRB systems. Solid lines are calculated with ${\rm DM}_{\rm tot}$, while dashed lines are calculated with the estimated ${\rm DM}_{\rm IGM}$. The red lines are for GRB 101011A, while green lines are for GRB 100704A. The black solid line regime corresponds to the constrained range of $\Omega_b f_{\rm IGM}$. The horizontal black dotted lines are the favored region of the Planck+WMAP results \citep{planckcollaboration13} corrected for $f_{\rm IGM} = 0.83$ derived from \cite{fukugita98}. The intersection of this band with the four lines give estimates of the redshifts of the two FRB/GRB systems: $z=(0.554, 0.687)$ for GRB 101011A, and $z=(0.130,0.246)$ for GRB 100704A.}
\label{fig:Omega_b_z}
\end{figure}

In Fig.\ref{fig:Omega_b_z} we present the constraints on $\Omega_b f_{\rm IGM}$ for the two FRB/GRB systems. For each case, we plot two lines: a solid line using ${\rm DM}_{\rm tot}$ and a dashed line using estimated ${\rm DM}_{\rm IGM}$. The lower limit on $z$ derived from the Amati relation requirement imposes an upper limit on $\Omega_b f_{\rm IGM}$. This upper limit is 0.101/0.114 for GRB 101011A, and 0.034/0.058 for GRB 100704A. Even though not tight, it is generally consistent with other measurements \citep{walker91,fukugita98,WMAP9yrs,planckcollaboration13}, and suggest that the matter component of the universe is dominated by dark matter. Such a consistency also supports that the two putative FRB-GRB associations reported by \cite{bannister12} are likely real.

One can also reverse the procedure to estimate $z$ of the two FRB/GRB systems using the available $\Omega_b f_{\rm IGM}$ constraints. Based on the $2\sigma$ best fit value of $\Omega_b = (0.046,0.052)$ from Planck+WMAP results \citep{planckcollaboration13} and the constraint on $f_{\rm IGM} \sim 0.83$ from baryon mass summation \citep{fukugita98}, one gets the $2\sigma$ range of $\Omega_b f_{\rm IGM}$: $(0.038,0.043)$ (horizontal lines in Fig.\ref{fig:Omega_b_z}). This gives estimated redshifts of the two FRB/GRB systems: $z=(0.554, 0.687)$ for GRB 101011A, and $z=(0.130,0.246)$ for GRB 100704A.

In the future, if $z$ is measured, one may also use the value of $\Omega_b f_{\rm
IGM}$ derived from our method along with the $\Omega_b$ value derived from the
standard method (CMB) to constrain $f_{\rm IGM}$.

\section{Constrain reionization history of the universe}

In the future, if FRB/GRB associations are commonly detected thanks to rapid follow-up observations of GRBs in the radio band, one would be able to well constrain the average $\Omega_b f_{\rm IGM}$ of the nearby universe. At $z>2$, $\chi_{\rm e,He}(z)$ would become $<1$, and at $z>6$, $\chi_{\rm e,H}(z)$ would also become $<1$. By measuring $z$ and ${\rm DM}_{\rm IGM}$ of FRB/GRB systems at these high redshifts, one would be able to constrain the reionization history of He and H in the universe based on Eq.(\ref{eq:DM_IGM}). The feasibility of this approach will be studied in a separate work.

\section{Summary and discussion}

In this Letter, we discuss some important cosmological implications of possible FRB/GRB associations. By measuring $z$ from the GRB and ${\rm DM}$ from the FRB, one can directly measure $\Omega_b f_{\rm IGM}$ using FRB/GRB systems at low redshifts. Even though no redshift measurements are available for GRB 101011A and GRB 100704A, we demonstrated that the method is applicable, and the derived loose constraints on $\Omega_b f_{\rm IGM}$ are consistent with results of other methods. This raises the prospects of mapping reionization history of the universe using FRB/GRB systems at higher redshifts.

The uncertainties of the method lie in precise determinations of other terms in ${\rm DM}_{\rm tot}$ (Eq.(\ref{eq:DM})). While ${\rm DM}_{\rm MW}$ can be more reliably constrained, ${\rm DM}_{\rm host}$ and ${\rm DM}_{\rm GRB}$ cannot. One can make an argument that both are relatively small values. If occasionally abnormally large ${\rm DM}_{\rm tot}$ is measured, the system can be excluded for cosmological studies, but could be used to study host galaxy properties (e.g. whether there exists a dense electronic disk or the host galaxy is near edge on) or the circumburst medium of the GRB (e.g. an over-dense wind environment).

Another issue is that the $\Omega_b f_{\rm IGM}$ measured for different lines of sights may fluctuate, and the scattering effect would introduce biases in FRB-GRB sample selection \citep{mcquinn14}.  Studying a large sample of FRB/GRB systems over a wide redshift range can give a more reliable averaged value of $\Omega_b f_{\rm IGM}$.

Within the FRB/GRB association picture, most FRBs are not supposed to be associated with GRBs, and their counterparts in other wavelengths may be faint \citep{zhang14}. If, on the other hand, the redshifts of these FRBs can be determined by other means, the methodology developed here can be also applied to those systems.

\medskip
We thank Zheng Zheng for helpful discussion and a referee for helpful comments. This work is partially supported by NASA under grant NNX10AD48G.





\end{document}